%% file: ManuscriptAccepted.tex
\documentclass[12pt,english]{article}
\usepackage[T1]{fontenc}
\usepackage[latin9]{inputenc}
\usepackage{geometry}
\usepackage{float}
\usepackage{amsmath}
\usepackage{amssymb}
\usepackage{graphicx}
\usepackage{setspace}
\PassOptionsToPackage{normalem}{ulem}
\usepackage{ulem}
\makeatletter

\providecommand{\tabularnewline}{\\}

\usepackage{lmodern}

\usepackage{amsmath}
\usepackage{graphicx}
\usepackage{setspace}
\usepackage{amssymb}
\usepackage{esint}
\newcommand{\tw}{\tilde{w}}
\newcommand{\ts}{\tilde{s}}

\newcommand{\vts}{v_{\ts}}

\newcommand{\zts}{z_{\ts}}

\newcommand{\prob}[1]{Pr \left\{#1\right\}}

\newcommand{\mean}[1]{E \left\{ #1 \right\}}

\newcommand{\be}{\begin{equation}}
\newcommand{\ee}{\end{equation}}

\newcommand{\dd}[1]{\, \mathrm{d} #1 \,}

\newcommand{\fin}{F^{-1}}
\newcommand{\gin}{G^{-1}}
\newcommand{\norm}[2]{\mathcal{N}(#1,#2)}
\newcommand{\hu}{\hat{u}}

\newcommand{\aw}{a_{\tilde{w}}}
\newcommand{\as}{a_{\tilde{s}}}
\newcommand{\ra}{r(\aw,\as)}
\newcommand{\rv}{r(v_w,v_s)}

\newcommand{\ta}{\tilde{a}}
\newcommand{\tv}{\tilde{v}}
\newcommand{\tz}{\tilde{z}}
\newcommand{\tg}{\tilde{g}}
\newcommand{\tG}{\tilde{G}}

\newcommand{\mnkn}[4]{\mean{\norm{#1}{#2}_{(#3,#4)}}}
\newcommand{\phia}[2]{\Phi^{-1}\left(\frac{#1}{#2}\right)}

\providecommand{\tabularnewline}{\\}

\usepackage{babel}
\usepackage[authoryear]{natbib}

\date{}

\makeatother

\usepackage{babel}
\begin{document}
\begin{center}
\textit{INCORPORATING HIDDEN COSTS OF ANNOYING ADS IN DISPLAY AUCTIONS}
\par\end{center}

\begin{center}
Valeria Stourm and Eric Bax\footnote{Valeria Stourm: Assistant Professor of Marketing, HEC Paris, stourmv@hec.fr,
Eric Bax: Yahoo Labs, ebax@yahoo-inc.com\inputencoding{latin1}{.}\inputencoding{latin9}
The authors are grateful for helpful comments and suggestions from
the review team, Michael Schwarz, Susan Athey, Preston McAfee, Ken
Wilbur, Pinar Yildirim, Eric Bradlow, Ron Berman, and participants
of our research seminar at Wharton. The authors especially thank John
Ledyard for advice about Pigovian auctions in practice. }
\par\end{center}
\begin{abstract}
Media publisher platforms often face an effectiveness-nuisance tradeoff:
more annoying ads can be more effective for some advertisers because
of their ability to attract attention, but after attracting viewers'
attention, their nuisance to viewers can decrease engagement with
the platform over time. With the rise of mobile technology and ad
blockers, many platforms are becoming increasingly concerned about
how to improve monetization through digital ads while improving viewer
experience. 

We study an online ad auction mechanism that incorporates a charge
for ad impact on user experience as a criterion for ad selection and
pricing. Like a Pigovian tax, the charge causes advertisers to internalize
the hidden cost of foregone future platform revenue due to ad impact
on user experience. Over time, the mechanism provides an incentive
for advertisers to develop ads that are effective while offering viewers
a more pleasant experience. We show that adopting the mechanism can
simultaneously benefit the publisher, advertisers, and viewers, even
in the short term. 

Incorporating a charge for ad impact can increase expected advertiser
profits if enough advertisers compete. A stronger effectiveness-nuisance
tradeoff, meaning that ad effectiveness is more strongly associated
with negative impact on user experience, increases the amount of competition
required for the mechanism to benefit advertisers. The findings suggest
that the mechanism can benefit the marketplace for ad slots that consistently
attract many advertisers. 

\thispagestyle{empty}

\newpage{}
\end{abstract}
\begin{center}
\textbf{1. INTRODUCTION}
\par\end{center}

Digital display ads can sometimes be a nuisance to viewers. Publisher
platforms, which charge to place ads on webpages, have a vested interest
in limiting negative ad impact on viewer experience because it jeopardizes
their long-term ability to deliver an engaged audience to advertisers
(Wilbur, Xu, and Kempe 2013). When estimating the advertising elasticity
of demand for TV, Wilbur et al. (2008) find that a 10\% increase in
advertising time leads to a 25\% reduction in audience size for a
popular broadcast network. Goldstein et al. (2014) experimentally
measure a platform's economic costs of serving annoying display ads
to viewers, and find that these costs seem to exceed the typical price
that digital platforms charge to advertisers. 

Despite their negative impact on viewer experience, distracting or
annoying ads tend to be noticed more and thus can be more valuable
to some advertisers. Ads that elicit more extreme, even negative,
reactions can sometimes be more effective than neutral ones because
they attract attention, facilitate memory for the advertised brand,
and may even enhance persuasion if their distraction inhibits counterarguing
(Aaker and Bruzonne 1985, Moore and Hutchinson 1983, Silk and Vavra
1974). As an example, more irritating television commercials are better
remembered and more likely to be recognized (Aaker and Bruzzone 1985).
Similarly, online gamers have explicit memory for ads that they find
annoying (Yeu et al. 2013). More recently, Zhou et al. (2016) examined
ads from Yahoo\textquoteright s mobile news stream. The quartile of
ads with the highest click-through rates had the highest proportion
of highly offensive ads. 

We use the term ad annoyance to refer to the degree to which an ad
irritates viewers. Aaker and Bruzzone (1985) describe annoying ads
as \textquotedblleft provoking, causing displeasure and momentary
impatience.\textquotedblright{} Annoying ads create an \textit{effectiveness-nuisance
tradeoff}: they may be more effective for some advertisers, but they
are a nuisance for viewers. This tradeoff challenges media platforms
that thrive on advertising, because some advertisers may be willing
to pay more to show more annoying ads. For example, Wall Street analyst
Richard Greenfield noted that ``while larger/more prominent ads clearly
detract from the Facebook user experience, we believe they come with
higher CPMs and should help re-accelerate revenues'' (Booton 2012).
For mobile platforms such as smart phones and tablets, the stakes
are higher because smaller screens and lower bandwidths make it difficult
to show effective ads, and viewers are less tolerant of annoyance.
The proliferation of ad blockers, which pose an even greater threat
to platform profits, is also growing as consumers become less tolerant
of annoying ads. 

A platform can partially manage the effectiveness-nuisance tradeoff
by limiting annoyance through factors directly under its control,
such as ad format or the number of ads on a page. In 2013, Google
reduced the ad load on mobile search by more than 50\%. Although this
substantially reduced revenue in the short term, the improved viewer
experience led to an increase of up to 4\% in user click-through rates
within ten weeks (Hohnhold, O\textquoteright Brien, and Tang 2015). 

However, ad characteristics that the platform does not directly control,
such as ad images and text, also strongly determine differences in
annoyance between ads. This paper focuses on how platforms can limit
ad annoyance that is driven by ad characteristics under advertiser
control. Based on data from Yahoo\textquoteright s ad feedback tool,
characteristics that determine ad annoyance include how easy it is
to read and understand ad text, how reliable users deem information
in the ad to be, symmetry, and aesthetic appeal (Zhou et al. 2016). 

Currently, most display advertising platforms use prohibitions to
control ad annoyance. Leading platforms prohibit ads containing violent
or offensive subject matter, containing animations that last more
than some number of seconds, and so forth. Unfortunately, under prohibitions,
ads will tend to the allowed limits of annoyance if annoyance is associated
with effectiveness. Also, prohibitions can be difficult to enforce,
because some platforms have limited control over ad content (Lambrecht
et al. 2014).

Instead of prohibitions, if advertisers internalize the costs of ad
annoyance through a \textquotedblleft tax\textquotedblright{} for
negative ad impact on viewer experience, then advertisers have an
incentive to develop ads that are less annoying as well as more effective.
Abrams and Schwarz (2008) proposed an auction that charges advertisers
for their ads\textquoteright{} impact on viewer experience. The charges
are like Pigovian taxes (Pigou 1912), because advertisers internalize
the negative externalities imposed by their ads. We refer to this
pricing mechanism, applied to a single display ad slot, as the Pigovian
second price auction (PSP). 

First, PSP subtracts from each advertiser's bid a charge for the (estimated)
reduction in long-term revenue to the platform that would be caused
by showing their ad. Next the mechanism conducts an auction based
on the adjusted bids. The winner is allocated ad space and pays the
adjusted-bid auction price plus the charge for the ad\textquoteright s
impact on user experience. 

Compared to the straightforward second-price auction (SP), PSP increases
the \textit{sum} of platform revenue and advertiser profit (Abrams
and Schwarz 2008). However, when ad effectiveness and nuisance are
positively associated, a scenario \uline{not} explicitly considered
by Abrams and Schwartz, PSP may increase platform revenue while decreasing
advertiser profit. In other words, the distribution of profits can
be skewed. If switching to PSP lowers expected profits for advertisers,
then the platform may suffer in the long term, because advertisers
may shift more of their business to other platforms. 

This paper shows that switching from a second-price auction to PSP
can simultaneously improve revenue for advertisers and the platform
while improving viewer experience, under certain marketplace conditions.
We show that the bid adjustment for ad impact on user experience in
PSP can benefit advertisers by increasing dispersion among advertiser
bids, so that in PSP the difference between the winning and runner-up
adjusted bids is greater on average than the difference between the
winning and runner-up bids from SP. These gaps are the winning advertiser's
profits under PSP and SP, respectively. Simultaneously, viewers benefit
from a more enjoyable experience, and that enables the platform to
benefit from an increase in future revenue by improving future audience
engagement. 

A previous paper, by Balachander et al. (2009), also uses a change
in dispersion among bids to show that a change in auction mechanisms
tends to change advertiser prices. That paper examines a change in
second-price auction mechanism from ranking and pricing by bid times
clickthrough rate to ranking and pricing strictly by bid. They show
that bids tend to have less dispersion than bid times clickthrough
rates, under the assumption that bids and clickthrough rates are correlated.
Thus, second prices tend to be closer to first prices, increasing
profit for the platform by decreasing profit for advertisers. In contrast,
this paper examines changing the second-price auction mechanism to
incorporate a charge for ad impact on user experience. We show that
this change can benefit the platform and advertisers simultaneously.
While the platform benefits from an increase in future revenue by
offering users a more engaging experience, advertiser profits increase
because bids adjusted for ad impact on user experience have more dispersion
than bids alone.

Since the relationship between bids and bids adjusted for ad impact
on user experience depends on the strength of the effectiveness-nuisance
tradeoff, we explicitly model the tradeoff to study how it impacts
profitability for both advertisers and the platform. We model the
strength of the tradeoff by allowing the ``hidden cost'' of ad annoyance
to be associated with an advertiser\textquoteright s willingness to
pay for a slot. A positive association reflects that more annoying
ads tend to provide a higher value to advertisers. We then apply the
model to show that even in the presence of the effectiveness-nuisance
tradeoff, a platform can impose a charge for annoying ads without
compromising the profits earned for itself or by its winning advertisers. 

A key finding is that a higher number of advertisers competing for
ad space makes it more likely that the platform can sustainably charge
for ad annoyance (i.e., without compromising advertiser profits or
platform revenues). We derive a formula that relates the effectiveness-nuisance
tradeoff to the minimum number of advertisers needed for PSP to benefit
viewers, advertisers, and the platform. The formula shows that even
platforms facing a strong effectiveness-nuisance tradeoff can sustainably
implement PSP if they can attract enough advertisers. 

This article is organized as follows. Section 2 presents the payoffs
to each party under auction mechanisms with and without a Pigovian
charge for ad impact on viewer experience. Section 3 presents a theorem
showing that advertisers can benefit from the Pigovian charge if they
are numerous enough. Section 4 develops a model for the effectiveness-nuisance
tradeoff and illustrates the theorem with two examples of joint distributions
for ad effectiveness and nuisance. Section 5 shows that when PSP improves
advertiser profits, it also simultaneously benefits viewers and platform
profits. Section 6 analyzes the case in which ad annoyance and effectiveness
are negatively associated instead of positively associated. Section
7 explores how strategic advertisers choose ads based on auction type
and how ad choice affects the results from previous sections. Finally,
Section 8 concludes and proposes directions for future work. 

\begin{center}
\textbf{2. AUCTION MECHANISMS}
\par\end{center}

This section describes the payoffs for the second-price auction (SP)
and the Pigovian second-price (PSP) auction. These descriptions remove
some complexity found in actual auctions for online display ads, such
as reserve prices. The bids here are per-impression bids. 

In SP, advertisers are ranked by their bids. Let $n$ be the number
of bidders, and let $b_{1},...,b_{n}$ be their bids. Let $w$ be
the index of the highest bid, and let $s$ be the index of the second-highest
bid. SP allocates the advertising opportunity to advertiser $w$ and
charges this winner the second price: $b_{s}$. 

PSP incorporates measures of the hidden costs of ad annoyance into
the pricing and selection rules. Let $z_{1},...,z_{n}$ be measures
of the hidden costs imposed on the platform by showing the ad creatives
associated with bids $b_{1},...,b_{n}$. We assume that $z_{i}$ is
known to both advertiser $i$ and the platform. 

In practice, platforms can estimate $z_{i}$ through a combination
of experimentation and predictive models of how viewership changes
over time as a function of ad features, viewer engagement metrics,
and editorial assessment. See Goldstein et al. (2014) for an example
on how hidden costs (i.e., a platform's economic costs of serving
annoying display ads to viewers) can be experimentally measured. Similarly,
Hohnhold, O'Brien and Tang (2015) developed a model to predict long-term
user behavior based on short-term user satisfaction metrics. Platforms,
such as Yahoo, Hulu, Facebook and Twitter, often monitor ad impact
on user experience through a variety of methods, including text links
that collect ad feedback from users on the relevance and attractiveness
of ads (Rohrer and Boyd 2004). 

The PSP mechanism first adjusts offers by subtracting each ad's annoyance
costs from its bid: $b_{1}-z_{1},...,b_{n}-z_{n}$. Then it holds
a second-price auction using the adjusted offers. Let  $a_{i}=b_{i}-z_{i}$
denote adjusted offers. Let $\tilde{w}$ and $\tilde{s}$ be the indices
of the highest and second-highest adjusted offers, respectively. The
winner $\tilde{w}$ is charged the second-highest adjusted offer plus
the hidden cost imposed by the winner's ad: $a_{\tilde{s}}+z_{\tilde{w}}$. 

Under both auctions, advertisers bid their private values $v_{1},...,v_{n}$
for the opportunity to advertise. For SP, the auction is truthful
because it is a Vickrey auction (Vickrey 1961). For PSP, advertisers
bid $b(v_{i}-z_{i})+z_{i}$ in equilibrium, where the function $b(v_{i})$
denotes an equilibrium bid in a Vickrey-Clarke-Groves (VCG) auction
(Vickrey 1961, Clarke 1971, Groves 1973, Abrams and Schwartz 2008).
Since only one ad opportunity is auctioned, $b(x)=x$. 

Any advertiser who has a negative valuation for showing their ad,
$v_{i}<0$, does not enter a bid in SP, because if they did, then
they would risk negative utility should they win and receive zero
utility if they lose. If an advertiser's adjusted bid is negative,
$v_{i}-z_{i}<0$, then PSP removes its adjusted bid from the auction.
Define $r(x_{1},x_{2})=\max(x_{1},0)-\max(x_{2},0)$. Then the profit
for the winning advertiser in SP is $\rv$, and the profit for the
winning advertiser in PSP (which may not be the same winner as for
SP) is $\ra$. This is because if $a_{\tilde{w}}>0$ and $a_{\tilde{s}}>0$,
then the PSP winner is charged $a_{\tilde{s}}+z_{\tilde{w}}$, leaving
a profit of 
\begin{equation}
v_{\tilde{w}}-a_{\tilde{s}}+z_{\tilde{w}}=(v_{\tilde{w}}-z_{\tilde{w}})-a_{\tilde{s}}=a_{\tilde{w}}-a_{\tilde{s}}.
\end{equation}

Table 1 reviews our notation and Table 2 summarizes the outcomes under
both auctions. The price, cost of annoyance, and platform profit formulas
assume $v_{w}$, $v_{s}$, $a_{\tilde{w}}$, and $a_{\tilde{s}}$
are all nonnegative. The last three outcomes in Table 2 summarize
when PSP benefits viewers ($z_{\tilde{w}}>z_{w}$), the platform ($v_{\tilde{s}}-z_{\tilde{s}}>v_{s}-z_{w}$),
and advertisers ($a_{\tilde{w}}-a_{\tilde{s}}>v_{w}-v_{s}$). 

\begin{table}[H]
\begin{onehalfspace}
\begin{centering}
\caption{Notation}
\par\end{centering}
\begin{centering}
\begin{tabular}{|c|c|}
\hline 
Notation & Definition\tabularnewline
\hline 
\hline 
$v_{i}$ & advertiser's private value\tabularnewline
\hline 
$z_{i}$ & cost of ad's annoyance to the platform\tabularnewline
\hline 
$a_{i}$ & adjusted bid: private value minus annoyance\tabularnewline
\hline 
$w,\:s$ & winner and runner-up indices in SP\tabularnewline
\hline 
$\tilde{w},\:\tilde{s}$ & winner and runner-up indices in PSP\tabularnewline
\hline 
\end{tabular}
\par\end{centering}
\end{onehalfspace}
\end{table}

\begin{table}[H]
\begin{doublespace}
\begin{centering}
\caption{Outcomes}
\par\end{centering}
\centering{}%
\begin{tabular}{|c|c|c|}
\hline 
Outcome & SP & PSP\tabularnewline
\hline 
\hline 
Basis for selection & $v_{i}$ & $v_{i}-z_{i}$\tabularnewline
\hline 
Value of slot to winner & $v_{w}$ & $v_{\tilde{w}}$\tabularnewline
\hline 
Price & $v_{s}$ & $v_{\tilde{s}}-z_{\tilde{s}}+z_{\tilde{w}}$\tabularnewline
\hline 
Cost of annoyance & $z_{w}$ & $z_{\tilde{w}}$\tabularnewline
\hline 
Profit for platform & $v_{s}-z_{w}$ & $v_{\tilde{s}}-z_{\tilde{s}}$\tabularnewline
\hline 
Profit for winning advertiser & $r(v_{w},v_{s})$ & $r(v_{\tilde{w}}-z_{\tilde{w}},v_{\tilde{s}}-z_{\tilde{s}})$\tabularnewline
\hline 
\end{tabular}
\end{doublespace}
\end{table}

\begin{figure}
\includegraphics[scale=0.7]{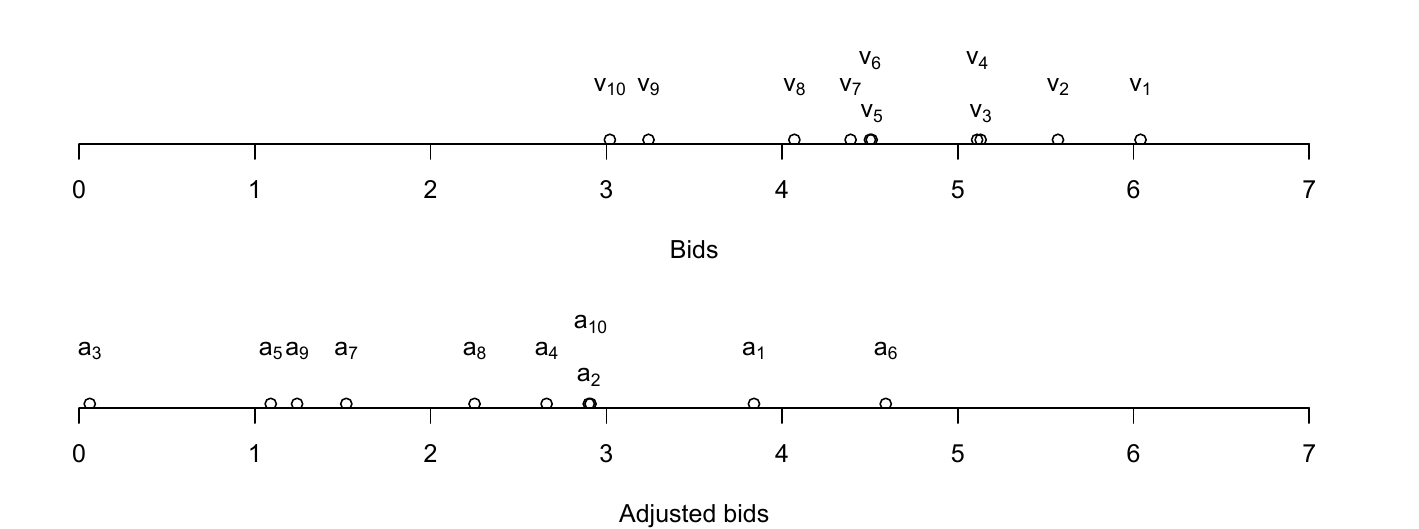}

\caption{A sample of bids and adjusted bids}

\label{fig: tenbids}
\end{figure}

We conclude this section with\textbf{ }a simple example that builds
intuition on how PSP can simultaneously benefit viewers, the platform,
and its advertisers. Consider ten advertisers with valuations $(v_{1},\ldots,v_{10})$
= (6.04, 5.57, 5.13, 5.11, 4.51, 4.50, 4.39, 4.07, 3.24, 3.02) and
ad impacts on user experience $(z_{1},\ldots,z_{10})$ = (2.20, 2.67,
5.07, 2.45, 3.42, -0.09, 2.87, 1.82, 2.00, 0.11). Their adjusted bids
$a_{i}=v_{i}-z_{i}$ are $(a_{1},\ldots,a_{10})$ = (3.84, 2.90, 0.06,
2.66, 1.09, 4.59, 1.52, 2.25, 1.24, 2.91). As shown in Figure \ref{fig: tenbids},
the respective winner and runner-up are advertisers one and two for
SP, and advertisers six and one for PSP. 

The data for this example was generated at random using a model we
will discuss in detail in the next section. The values $v_{i}$ were
drawn i.i.d. from a normal distribution with mean five and standard
deviation 1. The ad impacts on user experience $z_{i}$ were generated
by adding a quarter of $v_{i}$ to half of a value drawn i.i.d. from
a normal distribution with mean two and standard deviation 2. The
first term reflects an effectiveness-nuisance tradeoff, and the second
term reflects random variation in ad impact on user experience over
ads and advertisers. For ease of illustration, advertisers are ordered
by their bids $v_{i}$. 

Viewers, the advertisers as a group, and the platform simultaneously
benefit from PSP in this example. First, note that PSP clearly improves
viewer experience: $z_{6}>z_{1}$. The winning ad under SP (advertiser
1) has such a negative impact on user experience that showing it will
eventually decrease views enough for the publisher to lose $z_{1}=$
$2.20$. In contrast, the winning ad (advertiser 6) under PSP has
a positive impact on user experience: $z_{6}=-0.09$. 

Second, the profits for the platform are also greater for PSP: $a_{1}>v_{2}-z_{1}$.
For SP, the platform receives the second price $v_{2}$ = $5.57$
in immediate revenue from the winning advertiser, but incurs a $z_{1}=2.20$
eventual reduction in revenue due to the winning ad's impact on user
experience. Thus, the total platform revenue from SP is $5.57-2.20=3.37$.
For PSP, the platform receives from the winning advertiser the runner-up
adjusted bid of $a_{1}=$ $3.84$ minus a bonus of $z_{6}=-0.09$
for the winning ad's positive impact on user experience. The platform
eventually recovers the $0.09$ because the positive ad experience
drives more views and hence more revenue in the future. Thus, total
platform revenue for PSP is $3.84$. 

Third, the winning advertiser for PSP is more profitable than the
winning advertiser for SP. Notice that for both auctions, the winning
advertiser's profit is the difference between the highest and second-highest
values from a distribution: for SP, the distribution of offers $v_{i}$;
for PSP, the distribution of adjusted offers $v_{i}-z_{i}$. In this
example, the difference between the top two bids is $6.04-5.57=0.47$,
while the difference between the top two adjusted bids is $4.59-3.84=0.75$. 

The next section formally shows when PSP leads to greater advertiser
profits than SP. Notice from Figure \ref{fig: tenbids} that in this
example, adjusted bids are more dispersed than bids. This greater
variance in the distribution of adjusted bids is driven by random
variation in ad impact on user experience $z_{i}$. A consequence
of this dispersion is that PSP adjusted bids tend to have larger differences
between their top two adjusted bids than between the top two SP bids.
If this occurs, then advertiser profit for PSP is greater than for
SP, if the runner-up adjusted bid is also nonnegative. Notice also
from Figure \ref{fig: tenbids} that the distribution of adjusted
bids is shifted left from the distribution of bids. This occurs if
viewing ads is, on average over ads, a negative experience for viewers. 

An advertiser may have an adjusted bid greater than their bid (i.e.,
$a_{6}>v_{6}$) if their ad has such a positive impact on user experience
that it increases views for the publisher. These attractive ads are
more likely to be shown under PSP than under SP. It is also possible
for an advertiser with a positive bid $v_{i}$ to have a negative
adjusted bid. PSP excludes these advertisers, because their bid does
not cover the eventual cost to the publisher in reduced views from
showing the ad. In short, the Pigovian charge drives these ads out
of the auction. 
\begin{center}
\textbf{3. A THEOREM ON ADVERTISER PROFITS}
\par\end{center}

This section compares total expected profits for advertisers under
each auction.  The auctions may have different winning advertisers,
but in both cases the advertiser profit accrues to the winner. Recall
from Table 2 that advertiser profit equals the difference between
the top two offers under SP, and the difference between the top two
adjusted offers under PSP.  Also recall that $r(x_{1},x_{2})=\max(x_{1},0)-\max(x_{2},0)$.
Switching from SP to PSP increases the expected profits for advertisers
as a class when: 

\begin{equation}
E\{\ra\}>E\{\rv\}.\label{eq:adbase}
\end{equation}

In this section, we derive a theorem by comparing these expected profits
for a general set of distributions for bids $v_{i}$ and for adjusted
bids $a_{i}=v_{i}-z_{i}$. In the next section, we illustrate the
theorem using two examples of specific distributions. Let $f$ be
the pdf of bids $v_{i}$, and let $F$ be the cdf. Then the advertiser
surplus for SP can be written:
\begin{equation}
v_{w}-v_{s}=F^{-1}(F(v_{w}))-F^{-1}(F(v_{s})).\label{eq:Geq-2}
\end{equation}

If $f$ is continuous, then each $F(v_{i})$ is uniformly distributed
over the interval from zero to one: $U[0,1]$, under the probability
integral transform (David and Nagaraja 2003). The random variable
$F(v_{w})$ is the $n^{th}$ order statistic of $n$ uniform random
variables: $F(v_{1}),...,F(v_{n})$. This $n^{th}$ order statistic
is distributed $Beta(n,1)$ (David and Nagaraja 2003), so it has mean
$\frac{n}{n+1}$. Similarly, $F(v_{s})$ is the $n-1^{st}$ order
statistic, with distribution $Beta(n-1,2)$, and hence its mean is
$\frac{n-1}{n+1}$. To approximate the mean of the LHS of Equation
\ref{eq:Geq-2}, use the RHS and substitute the means of $F(v_{w})$
and $F(v_{s})$ for their values. Substituting these means inside
a non-linear function leads to an approximation of the LHS that becomes
more accurate as $n$ increases because the beta distributions become
more concentrated around their means\footnote{Our proof does not rely on this approximation; we use it only to simplify
this explanation.}. 
\begin{equation}
E\{v-v_{s}\}\approx F^{-1}\left(\frac{n}{n+1}\right)-F^{-1}\left(\frac{n-1}{n+1}\right).\label{eq:Geq-1-1}
\end{equation}

To analyze advertiser surplus for PSP, let $g$ be the pdf of adjusted
bids $a_{i}=v_{i}-z_{i}$ and let $G$ be the cdf. The expected advertiser
surplus can be approximated: 

\begin{equation}
E\{a_{\tilde{w}}-a_{\tilde{s}}\}\approx G^{-1}\left(\frac{n}{n+1}\right)-G^{-1}\left(\frac{n-1}{n+1}\right).\label{eq:Feq}
\end{equation}

Equations \ref{eq:Geq-1-1} and \ref{eq:Feq} show that advertiser
surplus is approximately equal to the difference between $F^{-1}\left(\frac{n-1}{n+1}\right)$
and $F^{-1}\left(\frac{n}{n+1}\right)$ for SP, and approximately
equal to the difference between $G^{-1}\left(\frac{n-1}{n+1}\right)$
and $G^{-1}\left(\frac{n}{n+1}\right)$ for PSP. Consider the following
approximation to Inequality \ref{eq:adbase} :
\begin{equation}
F^{-1}\left(\frac{n}{n+1}\right)-F^{-1}\left(\frac{n-1}{n+1}\right)<G^{-1}\left(\frac{n}{n+1}\right)-G^{-1}\left(\frac{n-1}{n+1}\right).
\end{equation}

\begin{figure}[H]
\includegraphics[scale=0.7]{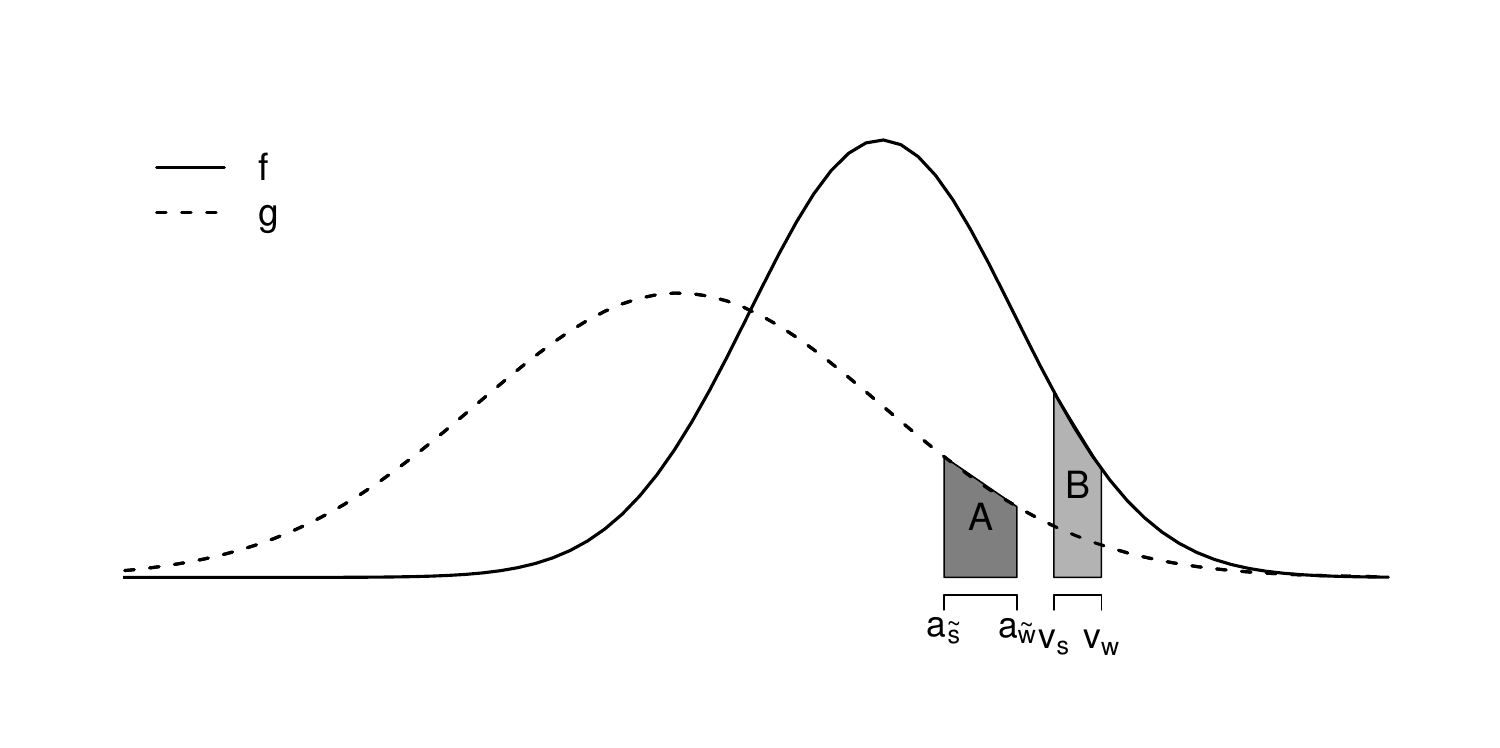}

\caption{Advertiser profits under each auction}

\label{fig: fgplot}
\end{figure}

Since F is a cdf, the area under the pdf $f$ from $F^{-1}\left(\frac{n-1}{n+1}\right)$
to $F^{-1}\left(\frac{n}{n+1}\right)$ must equal $\frac{1}{n+1}$.
Equivalently, after dividing the support of $f$ into $n+1$ segments,
each covered by $\frac{1}{n+1}$ of the area under $f$, advertiser
surplus is the distance covered by the second-to-last segment. Similarly,
the area under $g$ from $G^{-1}\left(\frac{n-1}{n+1}\right)$ to
$G^{-1}\left(\frac{n}{n+1}\right)$ must also equal $\frac{1}{n+1}$.
These are illustrated as areas ``A'' and ``B'' in Figure \ref{fig: fgplot}
for two hypothetical distributions $g$ and $f$.

Since A and B have equal areas (equal to $\frac{1}{n+1}$), when the
pdf $f$ is higher over B than the height of $g$ over A, the length
of the base (i.e., advertiser surplus) must be larger for A than for
B, making PSP more profitable than SP. In summary, if the second-to-last
segment supporting $\frac{1}{n+1}$ of the area under $g$ covers
more distance than the second-to-last segment supporting $\frac{1}{n+1}$
of the area under $f$, then PSP is favorable for advertisers. \textit{Equivalently,
a lower right tail in $g$ than in $f$ makes PSP favorable, given
sufficient numbers of competing advertisers to place $F^{-1}(\frac{n}{n+1})$,
$F^{-1}(\frac{n-1}{n+1})$, $G^{-1}(\frac{n}{n+1})$, and $G^{-1}(\frac{n-1}{n+1})$
out in the right tails of $f$ and $g$.}\textbf{ }We call this the
dispersion condition.

The dispersion condition is less likely to hold when there is a strong
effectiveness-nuisance tradeoff in the marketplace. If more annoying
ads are more effective for advertisers, then $v_{i}$ and $z_{i}$
are positively associated. An increase in the association between
$v_{i}$ and $z_{i}$ reduces the variance of adjusted bids, and makes
the right tail of $g$ higher. In the extreme case when $v_{i}$ and
$z_{i}$ are directly proportional, PSP harms advertisers. In this
case, adjusted bids $v_{i}-z_{i}$ are simply scaled versions of bids
$v_{i}$; the ranking of the bidders is the same under each auction,
and the same advertiser wins under each. However, the winner under
PSP is charged more, because they are charged for the annoyance of
their ad. 

If $v_{i}$ and $z_{i}$ are strongly (but not perfectly) positively
correlated, then the dispersion condition can be met if there are
enough advertisers participating in the auction. Theorem 1 (proven
in Appendix A) shows that when the dispersion condition is met, there
exists a minimum number of bidding advertisers such that the expected
profits for the winner are higher under PSP than under SP. 

\textbf{Theorem 1 }\textit{Let SP bids $v_{i}$ have a continuous
pdf $f$ and cdf $F$. Let PSP adjusted bids $a_{i}=v_{i}-z_{i}$
have a continuous pdf $g$ and cdf $G$. Suppose $g$ has a lower
right tail than $f$, in the sense that  $\exists\hat{u}<1$ such
that $\forall u>\hat{u}\::\:g(G^{-1}(u))<f(F^{-1}(u))$, $G^{-1}(\hat{u})\ge0$,
and $F^{-1}(\hat{u})<\infty$. Then $\exists\hat{n}(F,G)<\infty$
such that $\forall n\ge\hat{n}(F,G)$ the expected sum of advertisers'
profits from PSP is greater than that for SP: $E\{\ra\}>E\{\rv\}$.}

Now we briefly discuss the long-term implications of implementing
PSP, which are discussed in more detail in Section 7. Since Theorem
1 imposes minimal assumptions on the joint distribution of advertiser
valuations $v_{i}$ and nuisance costs $z_{i}$, it can be applied
to evaluate scenarios in which this joint distribution is expected
to change. The effectiveness-nuisance tradeoff in the marketplace
(which determines the joint distribution) is not expected to remain
constant over time, but can instead decrease in magnitude because
PSP provides advertisers with incentives to invest in developing ads
that are effective without being as annoying. As advertisers adapt
to PSP in the long run, the strength of the tradeoff should diminish,
making the dispersion condition more likely to hold. Visually, when
the correlation between $v_{i}$ and $z_{i}$ decreases, advertiser
profits under PSP (i.e., the base of area A in Figure \ref{fig: fgplot})
increase because the right tail of $g$ becomes lower while the area
over A still equals $\frac{1}{n+1}$. 

To summarize, since PSP provides advertisers with incentives to develop
better ads, advertiser profits under PSP are expected to increase
over time. Nevertheless, the platform must ensure that initially (for
the given current level of the effectiveness-nuisance tradeoff) PSP
can benefit its advertisers relative to SP, since platforms compete
with each other for advertisers. Thus, the next section analyzes the
implications of implementing PSP for a given level of the effectiveness-nuisance
tradeoff. 

\noindent \begin{center}
\textbf{4. A MODEL FOR THE EFFECTIVENESS-NUISANCE TRADEOFF}
\par\end{center}

To illustrate Theorem 1, we apply it to a marketplace with an effectiveness-nuisance
tradeoff in which advertisers with higher bids also tend to be those
with ads that are more annoying. Let an ad's hidden costs of annoyance
$z_{i}$ be a weighted function of scaled bids ($c\ge0$) and an independent
random variable $y_{i}$: 
\begin{equation}
z_{i}=\theta cv_{i}+(1-\theta)y_{i}\label{eq:eneq}
\end{equation}

The weight $\theta\in(0,1)$, which controls the positive association
between $v_{i}$ and $z_{i}$, is a measure of the effectiveness-nuisance
tradeoff present in the marketplace. As $\theta$ approaches 1, advertisers
with the highest bids tend to have ads with the most annoyance. As
$\theta$ approaches zero, $v_{i}$ and $z_{i}$ become independent,
so the tradeoff is not present. 

Under this model, the distributions for $v_{i}$ and $y_{i}$ jointly
define $f$ (the pdf of bids) and $g$ (the pdf of adjusted bids).
Note that adjusted bids under this model equal
\begin{equation}
a_{i}=v_{i}-z_{i}=(1-c\theta)v_{i}-(1-\theta)y_{i}.\label{eq:adjbid}
\end{equation}
Next we apply Theorem 1 to two examples of this model. In Example
1, both $v_{i}$ and $y_{i}$ are uniformly distributed. In Example
2, both are normally distributed. 

\textbf{4.1 Example 1: $v_{i}$ and $y_{i}$ uniformly distributed}

If both $v_{i}$ and $y_{i}$ have uniform distributions over the
interval from zero to one, then the area under $f$ is a square with
height one over {[}0,1{]}, and the area under $g$ is a trapezoid,
as shown in Figure \ref{fig: trapezoid}. It is the convolution of
two uniform distributions. The maximum height of the trapezoid is
$\frac{1}{1-c\theta}$, which is greater than one. 

If the second-highest adjusted bid is far enough along the right side
of the trapezoid that $g(a_{\tilde{s}})\le1$, then PSP is favorable
to SP for advertisers as a class. Visually, where the height of the
uniform $f$ is larger than the height of the trapezoid $g$ over
the two largest adjusted bids, the difference between the two largest
adjusted bids must be larger in expectation than the difference between
the two largest bids. Equivalently, if $g(x)\le1$ for $x\in[a_{\tilde{s}},a_{\tilde{w}}],$
then $g(x)\le f(x)$ over this domain, making PSP favorable to SP. 
\begin{center}
\begin{figure}[H]
$ $\includegraphics[scale=0.8]{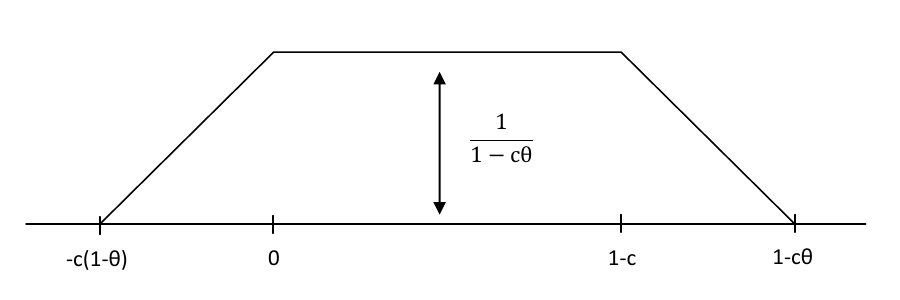}
\centering{}\caption{Density $g$ when $v_{i}$ and $y_{i}$ are correlated and uniformly
distributed }
\label{fig: trapezoid}
\end{figure}
\par\end{center}

This example highlights why, as more advertisers participate, PSP
is more likely to improve advertiser profits relative to SP. Recall
areas A and B from Figure \ref{fig: fgplot}. Since the areas of A
and B are equal, the area with the lower height must also have a longer
base. The length of the base is equal to the profits made by the winning
advertiser. 

In this example, bids have a uniform distribution $f$ in the shape
of a square with a height equal to 1, while adjusted bids have a distribution
$g$ in the shape of a trapezoid (with a height greater than 1). If
few advertisers participate, then it is likely that the top two adjusted
bids are near the middle of the trapezoid (where the height of $g$
is greater than 1). As more advertisers participate, the top two adjusted
bids are more likely to fall further out into the right tail of the
trapezoid, where the height of the pdf $g$ is less than one (i.e.,
the zone in which PSP leads to greater advertiser profits than SP). 

Next we calculate how many participating advertisers $n$ are needed
at a minimum for the expected second-highest adjusted bid to be far
enough along the right side of the trapezoid so that $g(a_{\tilde{s}})\le1$.
Using the approximation from Equation \ref{eq:Feq}, this condition
is approximately the same as
\[
g\left(G^{-1}\left(\frac{n-1}{n+1}\right)\right)<1.
\]

To solve for $n$, note that $s^{*}=(1-c\theta)\theta$ is the point
at which $g(s^{*})=1$. The above inequality holds when
\[
G^{-1}\left(\frac{n-1}{n+1}\right)\ge s^{*}
\]

or, equivalently,
\[
\frac{n-1}{n+1}\ge G(s^{*}).
\]

The area under $g$ from $s^{*}$ to the right end of the distribution
can be written as 1-$G(s^{*})$. This area is a triangle of height
1 and length $(1-c\theta)(1-\theta)$, so $G(s^{*})=1-\frac{1}{2}(1-c\theta)(1-\theta)$.
To find approximately how many advertisers are needed for PSP to be
favorable to SP for advertisers as a class, set this equal to $\frac{n-1}{n+1}$
and solve for $n$:
\begin{equation}
n\ge\frac{2}{(1-c\theta)(1-\theta)}-1.\label{eq:nineq}
\end{equation}

This lower bound increases as the effectiveness-nuisance tradeoff
increases (as  $\theta\rightarrow1$). As $\theta\rightarrow1$, the
sides of the trapezoid become steeper and cover less horizontal distance,
making the trapezoid resemble a rectangle. The next example calculates
a lower bound for $n$ when $v_{i}$ and $y_{i}$ are normally distributed.

\textbf{4.2 Example 2: $v_{i}$ and $y_{i}$ normally distributed}

Let $v_{i}$ and $y_{i}$ be normally distributed as shown in Equations
\ref{eq:vnorm} and \ref{eq:ynorm}.
\begin{eqnarray}
v_{i} & \sim & N(\mu_{v},\sigma_{v}^{2})\label{eq:vnorm}\\
y_{i} & \sim & N(\mu_{y},\sigma_{y}^{2})\label{eq:ynorm}
\end{eqnarray}

Then the distribution of bids $f$ is normal (Equation \ref{eq:vnorm}),
and the distribution of adjusted bids $g$ is also normal, with mean
$\mu_{a}=(1-c\theta)\mu_{v}-(1-\theta)\mu_{y}$ and variance $\sigma_{a}^{2}=(1-c\theta)^{2}\sigma_{v}^{2}+(1-\theta)^{2}\sigma_{y}^{2}$.
These two distributions are positively associated through the weight
$\theta$, which controls the strength of the effectiveness-nuisance
tradeoff. 

Using the normal model, we first show that Theorem 1 is satisfied
by $\hu=G(0)$ when the distribution of adjusted bids is more dispersed
than the distribution of bids: $\sigma_{a}>\sigma_{v}$. Then we use
$\hu=G(0)$ to estimate a lower bound on the number of advertisers
needed for PSP to benefit advertisers as a class when $\sigma_{a}>\sigma_{v}$.
We conclude by explaining how the magnitude of the effectiveness-nuisance
tradeoff parameter $\theta$ affects the condition $\sigma_{a}>\sigma_{v}$.

\textbf{4.2.1 Theorem 1 is satisfied by }$\hu=G(0)$\textbf{ when
$\sigma_{a}>\sigma_{v}$.}

Theorem 1 requires $\hu$ such that 1. $\forall u\geq\hu:g(\gin(u))<f(\fin(u))$,
2. $\gin(\hu)\geq0$ and 3. $\fin(\hu)<\infty$. For the first condition,
recall that any normal distribution cdf can be written in terms of
the standard normal cdf $\Phi$. For example,

\[
F(x)=\Phi\left(\frac{x-\mu_{v}}{\sigma_{v}}\right).
\]

Set $u=F(x)$ and $\fin(u)=x$. Then solve for $\fin(u)$:

\begin{equation}
\fin(u)=\mu_{v}+\Phi^{-1}(u)\sigma_{v}.\label{eq:foop}
\end{equation}

Recall that the pdf for a normal variable with mean $\mu_{v}$ and
standard deviation $\sigma_{v}$ is

\begin{equation}
f(x)=\frac{1}{\sqrt{2\pi}\sigma_{v}}e^{-\frac{1}{2}\left(\frac{x-\mu_{v}}{\sigma_{v}}\right)^{2}}.\label{eq:bloop}
\end{equation}

Substitute $x=\fin(u)$ and Equation $\ref{eq:foop}$ into Equation
$\ref{eq:bloop}$:

\[
f(\fin(u))=\frac{1}{\sqrt{2\pi}\sigma_{v}}e^{-\frac{1}{2}\left[\Phi^{-1}(u)\right]}.
\]

Similarly,
\[
g(\gin(u))=\frac{1}{\sqrt{2\pi}\sigma_{a}}e^{-\frac{1}{2}\left[\Phi^{-1}(u)\right]}.
\]

So 
\[
\forall u\in(0,1):g(\gin(u))=\frac{\sigma_{v}}{\sigma_{a}}f(\fin(u)).
\]

As a result, $g(\gin(u))<f(\fin(u))$ if and only if $\sigma_{a}>\sigma_{v}$,
and this holds for all $u\in(0,1)$. 

To meet the second condition, $\gin(\hu)\geq0$, set $\hu=G(0)$.
This also satisfies the third condition, $\fin(\hu)<\infty$, assuming
$\mu_{v}$, $\sigma_{v}$, $\mu_{a}$, and $\sigma_{a}$ are finite.
So for normal $f$ and $g$, Theorem 1 is satisfied by $\hu=G(0)$
if $\sigma_{a}>\sigma_{v}$. 

\textbf{4.2.2 Estimated lower bound on number of advertisers needed
when $\sigma_{a}>\sigma_{v}$}

Now we use $\hu=G(0)$ to analyze how many advertisers are needed
for PSP to benefit advertisers as a class when $\sigma_{a}>\sigma_{v}$.
Setting $\hu=G(0)$ means that PSP benefits advertisers when the runner-up
adjusted bid is nonnegative: $a_{\tilde{s}}\geq0$. Because $\mean{G(a_{\tilde{s}})}=\frac{n-1}{n+1}$,
the number of advertisers needed is about $n$ such that 
\[
G(0)=\frac{n-1}{n+1}.
\]

Since $G$ is the cdf of a normal distribution with mean $\mu_{a}$
and standard deviation $\sigma_{a}$, $G(0)\equiv\prob{g\leq0}=\Phi(-\frac{\mu_{a}}{\sigma_{a}})$.
So the number of advertisers required is about
\[
n=\frac{2}{1-\Phi(-\frac{\mu_{a}}{\sigma_{a}})}-1=\frac{2}{\Phi(\frac{\mu_{a}}{\sigma_{a}})}-1.
\]

We can use the standard normal cdf, with $\frac{\mu_{a}}{\sigma_{a}}$
as the Z-score, to find $\Phi(\frac{\mu_{a}}{\sigma_{a}})$. For example,
if $\frac{\mu_{a}}{\sigma_{a}}\approx-1$, then about 12 advertisers
are needed for the expected value of the runner-up adjusted bid to
be nonnegative:
\[
n=\frac{2}{\Phi(-1)}-1\approx12.
\]

In this case, the charge for ad annoyance makes average adjusted bids
negative, but the standard deviation of adjusted bids makes the probability
of each bid being nonnegative about 16\% (i.e., the probability of
a standard normal being at least one standard deviation above its
mean). For $\frac{\mu_{a}}{\sigma_{a}}\approx-2$, about 90 advertisers
are needed. In contrast, if $\frac{\mu_{a}}{\sigma_{a}}\approx0$,
then about half of adjusted bids are nonnegative, and three advertisers
are needed. 

To compare the magnitude of the figures from this example with actual
display auctions, we refer to summary statistics from a random sample
of second-price auctions from a leading online display advertising
platform, Microsoft Advertising Exchange. Celis et al. (2014) report
a summary distribution of the number of participating advertisers
through approximately 83,500 second-price auctions. The mean and median
number of advertisers per auction is 6, with a standard deviation
of 3, and a maximum of 15. In practice, the number of bidding advertisers
for each display slot can vary day-to-day across platforms and types
of slots. The ability to target consumers through information on their
preferences can reduce the number of advertisers that are interested
for a slot. Our findings suggest that platforms should implement the
Pigovian tax for competitive ad slots that consistently attract a
large group of bidders, rather than for ad slots with thinner markets. 

\textbf{4.2.3 The role of effectiveness-nuisance tradeoff parameter
$\theta$} 

Since $\sigma_{a}^{2}=(1-c\theta)^{2}\sigma_{v}^{2}+(1-\theta)^{2}\sigma_{y}^{2}$,
the inequality $\sigma_{a}>\sigma_{v}$, needed to satisfy Theorem
1, holds when
\begin{equation}
\sigma_{y}>\frac{\sqrt{1-(1-c\theta)^{2}}}{1-\theta}\sigma_{v}.\label{eq:foof}
\end{equation}

As $\theta$ approaches its upper limit, Inequality $\ref{eq:foof}$
does not hold, because setting $\theta=1$ makes the denominator zero.
Thus, PSP will not benefit advertisers when the costs of annoyance
are directly proportional to effectiveness: $z_{i}=cv_{i}$. The inequality
holds when there is no tradeoff, because setting $\theta=0$ makes
the numerator equal to zero. 

For intermediate cases, $0<\theta<1$, Inequality $\ref{eq:foof}$
holds if the independent component of $z_{i}$, which is $(1-\theta)y_{i}$,
boosts the standard deviation of adjusted bids more than the proportional
component, $c\theta v_{i}$, shrinks it. Once $\sigma_{a}$ becomes
larger than $\sigma_{v}$, PSP becomes more advantageous for advertisers
if there are a sufficient number of them to have nonnegative runner-up
and winning bids. 

\begin{center}
\textbf{5. VIEWER EXPERIENCE AND PLATFORM PROFIT}
\par\end{center}

This section shows how PSP benefits the platform and viewers. The
platform operates in a two-sided market: it offers content to attract
viewers, and it offers advertisers an opportunity to advertise to
the audience. The platform must ensure that a switch to PSP benefits
the entire marketplace: advertiser profits (discussed in previous
sections), viewer experience, and platform profits. 

\textbf{5.1 Viewer experience}

PSP is designed to improve viewer welfare because it selects advertisements
that lead to a better viewer experience. PSP improves (or at least
preserves) viewer experience on an auction-by-auction basis (Theorem
2), and it decreases the expected costs of ad annoyance (Theorem 3). 

\textbf{Theorem 2.} \textit{Switching from SP to PSP improves (or
at least preserves) viewer experience by reducing the costs of ad
annoyance: $z_{\tilde{w}}\le z_{w}$.}

\textit{Proof.} By the definition of $\tilde{w}$, $v_{w}-z_{w}\le v_{\tilde{w}}-z_{\tilde{w}}$,
which implies $z_{\tilde{w}}\le z_{w}-(v_{w}-v_{\tilde{w}})$. By
the definition of $w$, $v_{w}\ge v_{\tilde{w}}$. Hence the term
in parenthesis is positive, proving that $z_{w}\ge z_{\tilde{w}}.$
$\square$

\textbf{Theorem 3.} \textit{If ad annoyance costs to the platform
are not perfectly correlated with advertiser valuations ($\theta<1$),
then switching from SP to PSP decreases the hidden costs of ad annoyance
in expectation: $E\{z_{\tilde{w}}\}<E\{z_{w}\}.$}

\textit{Proof.} From Theorem 2, $z_{\tilde{w}}\le z_{w}$. Thus we
only need to show that $Pr\{z_{\tilde{w}}=z_{w}\}<1$. Since $z_{i}$
are drawn i.i.d. from some continuous distribution, $z_{i}$ values
for different advertisers $i$ have a zero probability of being equal:
$Pr\{z_{i}=z_{j}\}=0$ $\forall i\ne j$. Hence, we only need to show
that the auctions sometimes pick different winners: $Pr\{\tilde{w}=w\}<1$,
or equivalently, $Pr\{\tilde{w}\ne w\}>0$, which is true when $v_{i}$
and $z_{i}$ are not perfectly correlated. $\square$

\textbf{5.2 Platform profits}

Recall from Table 2 that the platform's profits from SP are $v_{s}-z_{w}$
and from PSP are $v_{\tilde{s}}-z_{\tilde{s}}$. PSP leads to greater
profits because it allows the platform to select better ads and to
charge advertisers for annoyance costs. In other words, it selects
ads based on total revenue for the platform $v_{i}-z_{i}$, while
SP selects ads only based on bids $v_{i}$.

Two extreme cases of the effectiveness-nuisance tradeoff, $\theta=1$
and $\theta=0$, provide further intuition. If $\theta=1$, the most
annoying ads are also the most effective, and $v_{i}$ and $z_{i}$
are perfectly correlated. To illustrate, suppose that $z_{i}=cv_{i}$
for a constant $c$$\ge0$. Both auctions produce the same ranking
of advertiser offers, but the platform's profits are greater under
PSP because it charges the winner a premium for its ad's impact on
user experience.

When there is no effectiveness-nuisance tradeoff ($\theta=0$), then
$z_{i}$ and $v_{i}$ are independent. In this case, PSP also improves
platform profits because the runner up in PSP is likely to have both
a high private value $v_{i}$ and low annoyance $z_{i}$. As the number
of bidding advertisers increases to infinity, the runner up almost
surely has one of the highest private values for the slot and one
of the least annoying ads. Consequently, PSP is likely to reduce annoyance
costs while maintaining a high ad price. 

The following theorem states that PSP produces at least as much expected
revenue for the platform as SP.

\textbf{Theorem 4. }\textit{Switching from SP to PSP is expected to
improve (or at least preserve) }platform profits: $\forall\theta\in[0,1]$,
$c\ge0$, and $n\ge3$, $E\{v_{\tilde{s}}-z_{\tilde{s}}\}\ge E\{v_{s}-z_{w}\}$.

See Appendix B for the formal proof. The theorem focuses on the normal
distribution example, but it is easy to see that it also applies to
the uniform distribution example and to other symmetric distributions.
The theorem assumes two or more nonnegative bids and adjusted bids.
If there is only one positive bid, then SP platform revenue is $-z_{w}$,
because it incurs the ad impact on user experience and shows the ad
for no charge.  If there is only one positive adjusted bid, then PSP
platform revenue is zero, because it charges only for the ad impact
on user experience. 

This section has shown that when there is an effectiveness-nuisance
tradeoff, PSP generally improves platform profits and viewer experience.
Previous sections showed that advertisers only benefit from PSP if
enough of them participate in the auction. So, in general, PSP can
simultaneously benefit the marketplace (viewer experience, advertiser
profit, and platform profit) when advertisers benefit as a class. 
\begin{center}
\textbf{6. ATTRACTIVE ADS}
\par\end{center}

Thus far, we have compared PSP to SP for models with an effectiveness-nuisance
tradeoff in which ads tend to have annoyance associated with the value
to the advertiser to show the ad. But this need not be the case. Some
platforms may not face the effectiveness-nuisance tradeoff and may
instead observe that advertisers with more attractive ads tend to
have a higher willingness to pay. 

This section considers the case in which annoyance is \textit{negatively
associated} with value. In this section, we will use the term \textit{attractiveness}
to mean the opposite of annoyance: attractiveness equals $-z_{i}$,
and we will say that an ad is attractive if $-z_{i}>0$, (i.e., if
the ad increases future views for the platform instead of imposing
hidden nuisance costs). 

The next two subsections show that under these circumstances PSP can
also benefit both advertisers and the platform, because it selects
ads that are both attractive and effective. The exception is the extreme
case in which attractiveness and willingness-to-pay are perfectly
(or very highly) correlated: then SP and PSP both select the most
attractive ad, but the same advertiser is compensated more under PSP
for the ad's attractiveness. 

\textbf{6.1 Advertisers benefit from attractive ads}

PSP gives attractive ads an advantage in the auction and a credit
based on attractiveness: the advertiser's bid $v_{i}$ is boosted
by $-z_{i}$ to form the adjusted bid. If that bid wins the auction,
then the advertiser is charged the runner-up adjusted bid minus the
$-z_{i}$ boost. If the advertiser with the greatest value for showing
their ad also has the most attractive ad, then PSP gives that advertiser
a price break compared to SP. In general, if attractiveness is independent
of or positively associated with value, then the winning advertiser
benefits on average, due to dispersion. To see this, alter the normal
model from Section 4.2 to have attractiveness (rather than annoyance)
positively associated with value, by replacing
\begin{equation}
z_{i}=\theta cv_{i}+(1-\theta)y_{i}
\end{equation}
with
\begin{equation}
z_{i}=-[\theta cv_{i}+(1-\theta)y_{i}].\label{eq:attractive}
\end{equation}
The term $\theta$ now controls the association between ad attractiveness
and value. Recall that adjusted bids are $a_{i}=v_{i}-z_{i}$, so
the mean and variance of the distribution of adjusted bids are
\begin{equation}
\mu_{a}=(1+c\theta)\mu_{v}+(1-\theta)\mu_{y}
\end{equation}
and
\begin{equation}
\sigma_{a}^{2}=(1+c\theta)^{2}\sigma_{v}^{2}+(1-\theta)^{2}\sigma_{y}^{2}.\label{beauty}
\end{equation}
From Subsection 4.2.1, Theorem 1 is satisfied by $\hu=G(0)$ if the
dispersion of PSP's adjusted bids is greater than the dispersion of
SP's bids: $\sigma_{a}>\sigma_{v}$. To see that this condition is
satisfied, examine the two terms on the right-hand side of Equation
\ref{beauty}. Note that the first term $(1+c\theta)^{2}\sigma_{v}^{2}>\sigma_{v}^{2}$
(because $c$ and $\theta$ are both greater than zero) and the second
term is non-negative ($(1-\theta)^{2}\sigma_{y}^{2}\geq0$), so $\sigma_{a}>\sigma_{v}$. 

\textbf{6.2 Platform profits and attractive ads}

Now consider whether the platform benefits from PSP if ad attractiveness
($-z_{i}$) is positively associated with the value to the advertiser
for showing the ad ($v_{i}$). Under SP the platform benefits from
the attractiveness of the winning ad without compensating the advertiser
for it. Under PSP, though, the platform compensates the winning advertiser
for the difference in attractiveness between the winning ad and the
runner-up ad. 

In the extreme case that attractiveness and value are perfectly correlated,
PSP harms platform profit because it selects the same attractive ad
but compensates the advertiser for its attractiveness. In this case,
$\tilde{s}=s$, so platform profit under PSP equals SP platform profit
minus the difference in attractiveness between the top two ads (recall
Table 2):
\begin{equation}
\vts-\zts=v_{s}-z_{w}+(z_{w}-z_{s}).
\end{equation}

However, if attractiveness is not directly proportional to value (i.e.,
the most highly-valued ad need not be the most attractive), then PSP
can benefit the platform because it may select an ad that offers the
publisher a greater sum of bid value and attractiveness than the ad
selected by SP. We illustrate this using the normal model from Section
4.2, with the change in sign for $z_{i}$ to have attractiveness (instead
of annoyance) associated with value (Equation \ref{eq:attractive}).
Specifically, we find an approximate condition for PSP to provide
the platform with greater expected profits than SP: $\mean{\vts-\zts}>\mean{v_{s}-z_{w}}$. 

First some notation. Let $\mnkn{\mu}{\sigma^{2}}{k}{n}$ be the expected
value of the $k$th greatest value among $n$ i.i.d. draws from the
distribution $\norm{\mu}{\sigma^{2}}$. Then, PSP platform profit
is
\begin{equation}
\mean{\vts-\zts}=\mnkn{(1+c\theta)\mu_{v}+(1-\theta)\mu_{y}}{(1+c\theta)^{2}\sigma_{v}^{2}+(1-\theta)^{2}\sigma_{y}^{2}}{2}{n}.
\end{equation}
Since $-z_{w}=\theta cv_{w}+(1-\theta)y_{w}$, and $y_{w}$ is independent
of the ranking by bid to select $w$, the inequality 
\begin{equation}
\mean{\vts-\zts}>\mean{v_{s}-z_{w}}
\end{equation}
is equivalent to 
\begin{align}
\mnkn{(1+c\theta)\mu_{v}+(1-\theta)\mu_{y}}{(1+c\theta)^{2}\sigma_{v}^{2}+(1-\theta)^{2}\sigma_{y}^{2}}{2}{n}\nonumber \\
>\mnkn{\mu_{v}}{\sigma_{v}^{2}}{2}{n}+c\theta\mnkn{\mu_{v}}{\sigma_{v}^{2}}{1}{n}+(1-\theta)\mu_{y}.
\end{align}

To approximate this inequality, approximate each expectation of an
order statistic by the inverse cdf of the expected cdf value, which
is $\frac{n-k+1}{n+1}$ for the $k$th order statistic out of $n$
samples (David and Nagaraja 2003). For our normal distributions, this
approximation is shown in Equation \ref{eq:approxOS}.
\begin{equation}
\mnkn{\mu}{\sigma^{2}}{k}{n}\approx\mu+\phia{n-k+1}{n+1}\sigma.\label{eq:approxOS}
\end{equation}
Under this approximation, the inequality becomes 
\begin{align}
(1+c\theta)\mu_{v}+(1-\theta)\mu_{y}+\phia{n-1}{n+1}\sqrt{(1+c\theta)^{2}\sigma_{v}^{2}+(1-\theta)^{2}\sigma_{y}^{2}}\\
>\mu_{v}+\phia{n-1}{n+1}\sigma_{v}+c\theta\mu_{v}+c\theta\phia{n}{n+1}\sigma_{v}+(1-\theta)\mu_{y}.\nonumber 
\end{align}

Cancel $(1+c\theta)\mu_{v}+(1-\theta)\mu_{y}$ from both sides and
combine terms:
\begin{equation}
\phia{n-1}{n+1}\sqrt{(1+c\theta)^{2}\sigma_{v}^{2}+(1-\theta)^{2}\sigma_{y}^{2}}>\left[\phia{n-1}{n+1}+c\theta\phia{n}{n+1}\right]\sigma_{v}.
\end{equation}

For $0<\theta<1$, the inequality holds if $\sigma_{y}$ is sufficiently
large:
\begin{equation}
\sigma_{y}^{2}>\frac{\left[1+c\theta\frac{\phia{n}{n+1}}{\phia{n-1}{n+1}}\right]^{2}-(1+c\theta)^{2}}{(1-\theta)^{2}}\sigma_{v}^{2}.
\end{equation}
As previously discussed, in the extreme case when $\theta=1$ (attractiveness
is directly proportional to value) the inequality does not hold. When
attractiveness is independent of value ($\theta=0$) the inequality
holds for any $\sigma_{y}>0$: 
\begin{equation}
\phia{n-1}{n+1}\sqrt{\sigma_{v}^{2}+\sigma_{y}^{2}}>\phia{n-1}{n+1}\sigma_{v}.
\end{equation}

\begin{center}
\textbf{7. STRATEGIC ADVERTISERS}
\par\end{center}

Advertisers tend to have multiple ads available to submit to online
advertising auctions. In this section, we consider how advertisers'
ability to choose an ad to submit based on whether the auction mechanism
is SP or PSP affects the results from the previous sections. Allowing
advertisers to be strategic increases the dispersion of PSP adjusted
bids, which reduces the minimum number of advertisers needed for the
Pigovian tax to increase advertiser profits. 

We first show that advertisers choose to submit their highest-value
ad for SP and their highest adjusted-value ad for PSP. Consequently,
we can apply results from the previous sections if we substitute the
distribution of values of the highest-value ads (per advertiser) for
the distribution of advertisers' values for ads, and we substitute
the distribution of adjusted bids of the highest adjusted-bid ads
(per advertiser) for the distribution of adjusted bids. Finally, we
discuss how the substituted distributions can vary from the distributions
used for the earlier results, affecting dispersion and hence advertiser
profitability, when advertisers choose ads based on the auction type. 

Let us begin with new notation that is summarized in Table 3. We will
continue to use $i$ to index advertisers, and we will also use $j$
to index an advertiser's ads. Let $v_{ij}$ be the value to advertiser
$i$ for showing their ad $j$. Let $z_{ij}$ be the ad's impact on
user experience, and let $a_{ij}=v_{ij}-z_{ij}$ be the ad's adjusted
bid. For an advertiser $i$, let $j^{*}$ index the ad with maximum
value and let $\tilde{j}$ index an ad with maximum adjusted bid. 

\begin{table}[h]
\begin{onehalfspace}
\begin{centering}
\caption{Notation for strategic advertisers}
\par\end{centering}
\centering{}%
\begin{tabular}{|c|c|}
\hline 
Notation & Definition\tabularnewline
\hline 
\hline 
$v_{ij}$ & advertiser $i$'s value for ad $j$\tabularnewline
\hline 
$j^{*}$ & index of max\{$v_{ij}$\} for advertiser $i$\tabularnewline
\hline 
$\tilde{j}$ & index of max\{$a_{ij}$\} for advertiser $i$\tabularnewline
\hline 
$v_{i}^{*},z_{i}^{*},a_{i}^{*}$ & characteristics of $j^{*}$ (ad that $i$ chooses for SP)\tabularnewline
\hline 
$\tilde{v}_{i},\tilde{\:z}_{i},\tilde{\:a}_{i}$ & characteristics of $\tilde{j}$ (ad that $i$ chooses for PSP)\tabularnewline
\hline 
\end{tabular}
\end{onehalfspace}
\end{table}

The following two lemmas outline how advertisers select which ad to
submit based on the auction type. We assume advertisers' ad values
and impacts on user experience are drawn at random. The lemmas allow
different distributions of ad values and impacts on user experience
for different advertisers.

\textbf{Lemma 1} \textit{For the SP auction, each advertiser maximizes
their expected profit by submitting a highest-value ad from the advertiser's
set of ads.} 

\textit{Proof. }This is a proof by contradiction. Suppose advertiser
$i$ submits an ad $j$ that has less than the maximum value among
advertiser $i$'s ads: $v_{ij}<v_{ij}^{*}$, and that maximizes their
expected profit. Instead, the advertiser could submit a maximum-value
ad and use the same bid as the advertiser used for ad $j$. Since
the advertiser uses the same bid, their probability of winning and
price if they win are the same. But if they win, then they receive
$v_{ij}^{*}-v_{ij}>0$ more value. $\square$

\textbf{Lemma 2} \textit{For the PSP auction, each advertiser maximizes
their expected profit by submitting a highest adjusted-bid ad from
the advertiser's set of ads, assuming the advertiser has an ad with
positive adjusted bid. }

\textit{Proof. }We will also prove this lemma by contradiction. Suppose
advertiser $i$ submits an ad $j$ that has $a_{ij}<a_{i\tilde{{j}}}$,
bids $b$, and that maximizes the advertiser's expected profit. Instead,
the advertiser could submit ad $\tilde{j}$ and bid $b-z_{ij}+z_{i\tilde{j}}$.
The advertiser would have the same adjusted bid, because $(b-z_{ij}+z_{i\tilde{j}})-z_{i\tilde{j}}=b-z_{ij}$.
So the advertiser would have the same probability of winning the auction.
Let $s$ be the adjusted bid for the second-place offer, or zero if
that adjusted bid is negative. If the advertiser wins with ad $j$,
then their profit is $v_{ij}-(s+z_{ij})=a_{ij}-s$. If the advertiser
won with ad $\tilde{j}$, then their profit would be $v_{i\tilde{j}}-(s+z_{i\tilde{j}})=a_{i\tilde{j}}-s$.
Since $a_{i\tilde{j}}>a_{ij}$, profit would be greater using ad $\tilde{j}$
than ad $j$. $\square$

We now prove a theorem that is similar to Theorem 1, but which applies
to strategic advertisers choosing from a set of ads. As in previous
sections, let $w$ and $s$ index the winning and second-place advertisers
for SP, and let $\tw$ and $\tilde{s}$ index the winning and second-place
advertisers for PSP. As shown in Table 3, let $v_{i}^{*}$, $z_{i}^{*}$,
and $a_{i}^{*}$ be the value, ad impact on user experience, and adjusted
bid ($a_{i}^{*}=v_{i}^{*}-z_{i}^{*}$) for the ad that advertiser
$i$ chooses for SP. Let $\tv_{i}$, $\tz_{i}$, and $\ta_{i}$ be
the value, ad impact on user experience, and adjusted bid ($\ta_{i}=\tv_{i}-\tz_{i}$)
for the ad that advertiser $i$ chooses for PSP. 

\textbf{Theorem 5} \textit{Let the distribution of $v_{i}^{*}$ have
have a continuous pdf $f^{*}$ and cdf $F^{*}$. Let the distribution
of $\ta_{i}$ have a continuous pdf $\tg$ and cdf $\tG$. Suppose
$\tg$ has a lower right tail than $f^{*}$, in the sense that $\exists\hu<1$
such that $\forall u>\hu$: $\hat{g}(\tilde{G}^{-1}(u))<f^{*}(F{}^{*-1}(u))$,
$\tilde{G}^{-1}(\hat{u})\geq0$, and $F^{*-1}(\hu)<\infty$. Then
$\exists\hat{n}(F^{*},\tilde{G})<\infty$ such that $\forall n\geq\hat{n}(F^{*},\tilde{G})$
the expected sum of advertisers' profits from PSP is greater than
for SP: $E\{r(\ta_{\tw},\ta_{\ts})\}>E\{r(v_{w}^{*},v{}_{s}^{*})\}$.}

\textit{Proof. }As discussed in Section 2, SP and PSP are truthful
auctions, so advertisers bid $v_{i}^{*}$ in SP and have adjusted
bids $\ta_{i}$ in PSP. The remainder of the proof is the proof of
Theorem 1, with $v_{i}^{*}$, $f^{*}$, $F^{*}$, $\ta_{i}$, $\tg$,
and $\tG$ substituted for $v_{i}$, $f$, $F$, $a_{i}$, $g$, and
$G$, respectively. $\square$

In general, if $\tg$ is more dispersed than $f^{*}$ and there is
a sufficient number of advertisers bidding, then PSP increases advertiser
profit over SP. Note that before deciding whether to implement PSP,
a platform does not observe the distribution of available advertisements,
and thus does not directly observe $\tg$. For comparison to the results
in earlier sections, we can treat $v_{i}^{*}$ as equivalent to $v_{i}$,
and we can consider $f^{*}=f$. However, in general, the distribution
of adjusted bids will not be the same with vs. without advertiser
choice: $\tg\not=g$. In other words, $\ta_{i}$ (the highest adjusted
bid for advertiser $i$) will not have the same distribution as $a_{i}$
(the adjusted bid of $i$'s highest value ad). Note that $a_{i}=a{}_{i}^{*}=v{}_{i}^{*}-z{}_{i}^{*}$.

In the model in Section 4, $z_{i}^{*}$ is allowed to have some dependence
on $v_{i}^{*}$. We showed that subtracting $z_{i}^{*}$ from $v_{i}^{*}$
produces adjusted bids with greater dispersion than the original bids
if there are enough bidders. If advertisers can choose a different
ad for PSP than the highest-bid ad they choose for SP, then instead
of variation in $a_{i}^{*}-v{}_{i}^{*}=-z{}_{i}^{*}$ without advertiser
choice, with advertiser choice, the results rely on variation in $\ta_{i}-v{}_{i}^{*}=(\ta_{i}-a{}_{i}^{*})-(a{}_{i}^{*}-v{}_{i}^{*})=-z{}_{i}^{*}+(\ta_{i}-a{}_{i}^{*})$.
The additional term $\ta_{i}-a{}_{i}^{*}$ is the maximum amount that
advertiser $i$ can increase their adjusted bid by switching to a
different ad for PSP than the one they would choose for SP. The term
cannot be negative, since the advertiser has the option to use the
same ad for both SP and PSP. 

Consider an auction with two bidders, advertisers 1 and 2. Suppose
advertiser 1 has value $v_{i}^{*}=1$ and advertiser 2 has value $v_{2}^{*}=1.20$.
They bid those values in SP, and advertiser wins with a $0.20$ profit.
Suppose their highest-value ads have equal impact on user experience:
$z_{1}^{*}=z_{2}^{*}=0.30$. For PSP without strategic ad choices,
the adjusted bids are $a_{1}^{*}=0.70$ and $a_{2}^{*}=0.90$, and
again advertiser 2 wins with a $0.20$ profit. 

Now consider the effects of strategic ad choices. Given some values
of $\tilde{a}_{2}$ and $\tilde{a}_{1}$ (the adjusted bids for the
ads chosen for PSP by advertisers two and one), if $\tilde{a}_{2}-a_{2}^{*}>\tilde{a}_{1}-a_{1}^{*}$,
then advertiser 2 increases their profit beyond $0.20$. Note that
$\tilde{a}_{i}-a_{i}^{*}=(\tilde{v}_{i}-v_{i}^{*})-(\tilde{z}_{i}-z_{i}^{*})$,
so $\tilde{a}_{2}-a_{2}^{*}$ represents advertiser 2's flexibility
to give up some value, $\tilde{v}_{2}-v_{2}^{*}<0$, to gain more
in the adjusted bid by improving user impact by $-(\tilde{z}_{2}-z_{2}^{*})>0$.
In general, if without strategic choice, PSP winning advertisers have
more of this flexibility than runner-up advertisers, then strategic
ad choice will increase their PSP profits. Also, if there are more
advertisers and more variation in this flexibility over advertisers,
then that variation will drive increased profits for strategic choice
in the same way that variation in ad impact drives increased profits
from SP to PSP -{}- by increasing dispersion. 

Given a joint distribution for $v_{i}^{*}$, $z_{i}^{*}$, and $\ta_{i}-a{}_{i}^{*}$,
we can build a model as we did in Section 4 for a joint distribution
over $v_{i}^{*}$ and $z_{i}^{*}$, determine the conditions for $\ta_{i}$
to have more dispersion than $v_{i}^{*}$, and apply Theorem 5 to
show that PSP benefits advertisers under those conditions. We will
start with the model from Section 4, identify $v_{i}^{*}$ with $v_{i}$
in that model, and consider how to alter the terms for $z_{i}^{*}$
so that they describe the distribution of $z_{i}^{*}-(\ta_{i}-a{}_{i}^{*})$
instead. Of course, if $\ta_{i}-a{}_{i}^{*}=0$ for all advertisers
(i.e., if advertisers' highest-value ads are also their highest adjusted-bid
ads), then the model from Section 4 needs no adjustments for advertiser
choice.

If we model the difference in adjusted values due to ad choice by
a constant: $\ta_{i}-a{}_{i}^{*}=d>0$, then the model in Equation
\ref{eq:adjbid} becomes
\[
\ta_{i}=v{}_{i}^{*}-z{}_{i}^{*}+(\ta_{i}-a{}_{i}^{*})=(1-c\theta)v_{i}-(1-\theta)y_{i}+d.
\]
The constant $d$ moves the distribution of $\ta_{i}$ to the right
compared to that of $a_{i}^{*}$ (i.e., $\tg$ is $g$ shifted to
the right). The distribution of $v_{i}^{*}$ the same as that of $v_{i}$,
so $f^{*}=f$. Under the shift, $\tg(\tG^{-1}(u))=g(G^{-1}(u))$,
so the lower right tail condition remains the same. 

For the normal version of the model, following the logic from Subsection
4.2.1, Theorem 5 is satisfied by $u=\tG(0)$ if $\sigma_{a}>\sigma_{v}$.
With a shift to the right, $\tG(0)<G(0)$, so \textit{fewer advertisers
(or the same number) are needed for PSP to benefit advertisers on
average.} In short, the gain in adjusted bids due to the ability to
choose among ads can only increase the number of advertisers with
positive adjusted bids, so it can only increase participation in PSP. 

Now suppose $\ta_{i}-a{}_{i}^{*}$ is random per advertiser: $\ta_{i}-a{}_{i}^{*}=d_{i}$.
The random variable $d_{i}$ must be bounded below by zero, since
$\ta_{i}\geq a{}_{i}^{*}$. For simplicity, assume the distribution
of $d_{i}$ has positive mean, is symmetric and smooth over its range,
and has a zero or negative second derivative over its range. For now,
assume $d_{i}$ is independent of both $v_{i}^{*}$ and $z_{i}^{*}$.
The model becomes 
\[
\ta_{i}=v{}_{i}^{*}-z{}_{i}^{*}+(\ta_{i}-a{}_{i}^{*})=(1-c\theta)v_{i}-(1-\theta)y_{i}+d_{i}.
\]
 Adding $d_{i}$ makes the distribution of $\ta_{i}$ flatter than
the distribution of $a_{i}^{*}$ in addition to shifting it to the
right. (The distribution of $\ta_{i}$ is the convolution of the distributions
of $a_{i}^{*}$ and $d_{i}$.) As a result, adjusted bids in PSP have
more dispersion when advertisers can select ads strategically.

When $d_{i}$ is not independent of $v_{i}^{*}$ and $z_{i}^{*}$,
dispersion will tend to increase if $d_{i}$ tends to increase with
$v_{i}^{*}$. This models a scenario in which choice allows the rich
to get richer: advertisers with higher highest-value ads can choose
alternative ads that improve ad impact more per unit decrease in value,
that give up more value in exchange for even more improvement in ad
impact, or both. Similarly, dispersion will tend to increase if $d_{i}$
tends to increase as $z_{i}^{*}$ increases. This corresponds to a
scenario in which advertisers who rely on ad nuisance for ad effectiveness
are less likely to be able to maintain ad effectiveness while decreasing
nuisance.

\begin{center}
\textbf{8. CONCLUSION}
\par\end{center}

This article examined an auction mechanism that charges advertisers
a Pigovian tax for the economic costs to the publisher of ad annoyance.
The Pigovian second price auction (PSP) adds a charge to the second
price (SP) auction for the costs of negative ad impact on viewer experience
to the publisher and selects among bids that are adjusted for this
charge. The charge is analogous to a Pigovian tax because  advertisers
internalize the costs of negative ad impact on viewer experience. 

While PSP can align advertiser and viewer incentives, a platform may
be hesitant to implement it because the Pigovian charge may decrease
the profits of its advertisers. But this not need be the case. This
article analyzed conditions for PSP to reduce ad annoyance without
reducing expected advertiser profits. 

We developed a general model of the effectiveness-nuisance tradeoff.
One advantage of the model is that it imposes minimal assumptions
on the joint distribution of advertiser valuations and nuisance costs.
Properties of order statistics were applied to compare the outcomes
of PSP and SP. 

We found that when enough advertisers are willing to bid for an ad
slot, PSP can improve the quality of ads without reducing expected
advertiser profits. When PSP benefits advertisers, then the platform's
own expected profits and the viewer experience also improve. We also
discussed how the model can assess long-term outcomes as advertisers
develop ads that are more effective and less annoying, and found that
when advertisers can strategically select ads in response to the auction
type, this reduces the number of advertisers required for PSP to improve
the marketplace.

We also showed how to estimate how many advertisers must bid for PSP
to simultaneously improve viewer experience, platform profits, and
advertiser profits. For example, when the underlying distributions
of the effectiveness-nuisance model are characterized through linked
normal distributions, as presented in Section 4.2, a Z-score statistic
can be used to estimate the minimum number of advertisers required
to implement PSP to the benefit of the marketplace. The Z-score varies
with advertiser valuations, ad nuisance costs, and the association
between them. Our findings suggest that platforms should implement
the Pigovian tax for competitive ad slots that consistently attract
a large group of bidders, rather than for ad slots with thinner markets. 

We outline two avenues for future research. The first is to examine
the implications of implementing PSP under platform competition. Since
we found that the advertisers of a single platform benefit from PSP
when there are enough bidders, suppose that only the most competitive
platforms initially adopt PSP, while smaller platforms with fewer
advertisers do not. Since PSP profitably improves viewer experience
and attracts the best advertisers, then, for the larger platforms,
switching to PSP would expand their lead over smaller competitors,
and advertisers would face greater incentives to improve their ads.\textbf{
}The marketplace dynamics of platform competition may also be influenced
by long-run objectives such as growing the viewer base or discouraging
viewers from adopting ad blocking technologies. 

The second avenue is to investigate the role of ad annoyance that
is controlled by platforms rather than advertisers. For example, some
platforms choose to sell ad formats that are obtrusive to viewers
but can sell at higher prices. In the long run, this strategy may
encourage more consumers to adopt ad-blocking technologies that pose
an even greater threat to platform profits than the nuisance costs
of annoying ads. 

We conclude by emphasizing the importance and urgency of researching
how to address the effectiveness-nuisance tradeoff, a challenge for
monetizing digital advertising. Managing this tradeoff is especially
critical in the growing industry of mobile display advertising, in
which ads can interfere with user experience because of lower bandwidths
and smaller screens. As Forbes magazine observes, ``users find mobile
ads invasive and annoying - even more so than desktop ads which are
easier to tune out. And as Facebook sees its user growth and engagement
slow (particularly with younger users looking elsewhere) it needs
to be careful not to do anything to hasten the exits\textquoteright \textquoteright{}
(Kosner 2012). Viewers in the mobile market have more leverage because
they drive app adoption and are more likely to generate content for
others to view. Similarly, hardware platforms are also sensitive to
negative ad impact, and have even banned some apps (or their advertisements)
based on negative impact on viewer experience. For example, Apple
initially prohibited all ads on some popular apps including YouTube
to encourage adoption of the iPad (Efrati 2012), and more recently
Apple released an iOS9 operating system that allows ad-blocking software. 

This study illustrated how to sustainably manage the effectiveness-nuisance
tradeoff of online display ads through a Pigovian pricing mechanism.
Promoting a better ad experience for users may be part of a sustainable
solution, avoiding a cat-and-mouse game of increasingly sophisticated
ad blockers chasing increasingly sophisticated ways to embed ads in
content or have ads masquerade as content. We hope our findings encourage
future work on how to improve monetization through digital display
ads in ways that improve the digital experience for viewers.

\bibliographystyle{plain}
\nocite{*}
\bibliography{pigovrefs_2016}
\input{proof2.tex} \input{appendixb_v2.tex}
\end{document}

%% file: proof2.tex
\noindent \begin{flushleft}
\textbf{APPENDIX A:} Proof of Theorem 1
\par\end{flushleft}

Let $r(x_1, x_2) = max(x_1, 0) - max(x_2, 0)$. Then the advertiser profit from PSP is $\ra$, and the advertiser profit from SP is $\rv$. To prove the theorem, we will show that
\be
\mean{\ra} - \mean{\rv} > 0.
\ee

To begin, we will transform $\mean{\rv}$ and $\mean{\ra}$ into integrals that are easy to compare. By definition,
\be
\mean{\rv} = \int_{x_1 = - \infty}^{\infty} \int_{x_2 = -\infty}^{x_1} r(x_1, x_2) q(x_1, x_2) \dd{x_1} \dd{x_2}, \label{defmrv}
\ee
where $q(x_1, x_2)$ is the joint pdf for the maximum ($x_1$) and runner up ($x_2$) among $n$ i.i.d. samples from a distribution having pdf $f$ and cdf $F$. Since there are ${{n}\choose{2}}$ ways to choose the top two samples and $2$ ways to choose which is the maximum, 
\be
q(x_1, x_2) = 2 {{n}\choose{2}} F(x_2)^{n-2} f(x_1) f(x_2).
\ee
Substitute into Equation \ref{defmrv}:
\be
\mean{\rv} = \int_{x_1 = - \infty}^{\infty} \int_{x_2 = -\infty}^{x_1} r(x_1, x_2) 2 {{n}\choose{2}} F(x_2)^{n-2} f(x_1) \dd{x_1} f(x_2) \dd{x_2}.
\ee
Use two changes of variables: $u_1 = F(x_1)$ and $u_2 = F(x_2)$. Since $\frac{\dd{u}}{\dd{x_1}} = f(x_1)$, $\dd{u} = f(x_1) \dd{x_1}$. Likewise, $\dd{u_2} = f(x_2) \dd{x_2}$. Then
\be
\mean{\rv} = \int_{u_1 = 0}^{1} \int_{u_2 = 0}^{u_1} r(\fin(u_1), \fin(u_2)) 2 {{n}\choose{2}} u_2^{n-2} \dd{u_1} \dd{u_2}.
\ee
Let 
\be
p(u_1,u_2) = 2 {{n}\choose{2}} u_2^{n-2}.
\ee
(This is the joint pdf for the maximum and runner up among $n$ i.i.d. samples from $U[0,1]$.) Then
\be
\mean{\rv} = \int_{u_1 = 0}^{1} \int_{u_2 = 0}^{u_1} r(\fin(u_1), \fin(u_2)) p(u_1, u_2) \dd{u_1} \dd{u_2}.
\ee
Applying the same process to $\mean{\ra}$, but with variable substitutions $u_1 = G(x_1)$ and $u_2 = G(x_2)$:
\be
\mean{\ra} = \int_{u_1 = 0}^{1} \int_{u_2 = 0}^{u_1} r(\gin(u_1), \gin(u_2)) p(u_1, u_2) \dd{u_1} \dd{u_2}.
\ee
Combine the integrals for $\mean{\rv}$ and $\mean{\ra}$:
\be
\mean{\ra} - \mean{\rv}
\ee
\be
= \int_{u_1 = 0}^{1} \int_{u_2 = 0}^{u_1} [r(\gin(u_1), \gin(u_2)) - r(\fin(u_1), \fin(u_2))] p(u_1, u_2) \dd{u_1} \dd{u_2}.
\ee
Split the integral based on whether $u_2 \geq \hu$:
\be
= \int_{u_2 = 0}^{\hu} \int_{u_1 = u_2}^{1} [r(\gin(u_1), \gin(u_2)) - r(\fin(u_1), \fin(u_2))] p(u_1, u_2) \dd{u_1} \dd{u_2}
\ee
\be
+ \int_{u_2 = \hu}^{1} \int_{u_1 = u_2}^{1} [r(\gin(u_1), \gin(u_2)) - r(\fin(u_1), \fin(u_2))] p(u_1, u_2) \dd{u_1} \dd{u_2}.
\ee

Call the first double integral I and the second double integral II. First, we will show that I can be made arbitrarily close to zero by selecting $n$ sufficiently large. Then we will show that II is positive.

Note that
\be
I \geq \prob{u_2 \leq \hu} \min_{u_2 \leq \hu, u_1 \geq u_2}\{r(\gin(u_1), \gin(u_2)) - r(\fin(u_1), \fin(u_2))\}.
\ee
Recall that $\mean{u_2} = \frac{n-1}{n+1}$, and $u_2 \in [0,1]$. By Markov's Inequality, 
\be
\prob{u_2 \leq \hu} = \prob{1 - u_2 \geq 1 - \hu} \leq \frac{\mean{1 - u_2}}{1 - \hu} = \frac{\frac{2}{n+1}}{1 - \hu}.
\ee
Since $\hu < 1$, $1 - \hu > 0$. So we can make $\prob{u_2 \leq \hu}$ arbitrarily close to zero by choosing $n$ sufficiently large. To show that we can make I arbitrarily close to zero, we also need to show that 
\be
\min_{u_2 \leq \hu, u_1 \geq u_2}\{r(\gin(u_1), \gin(u_2)) - r(\fin(u_1), \fin(u_2))\}
\ee
is finite. 

To do that, consider two cases: $u_1 \leq \hu$ and $u_1 > \hu$. Start with $u_1 \leq \hu$. First, $r(\gin(u_1), \gin(u_2)) \geq 0$, because the gap between  the maximum and runner-up bids is nonnegative. Second, $- r(\fin(u_1), \fin(u_2)) \geq - r(\fin(\hu), 0)$, since $u_1 \leq \hu$ and $\min(\fin(u_2),0) \geq 0$. So
\be
r(\gin(u_1), \gin(u_2)) - r(\fin(u_1), \fin(u_2)) \geq - r(\fin(\hu),0). 
\ee
If $\fin(\hu) \leq 0$, then the RHS is zero. If not, then it is $-\fin(\hu)$, which is finite because we assume $\fin(\hu) < \infty$.

For the case $u_1 > \hu$,
\be
r(\gin(u_1), \gin(u_2)) - r(\fin(u_1), \fin(u_2))
\ee
\be
= [r(\gin(\hu), \gin(u_2)) - r(\fin(\hu), \fin(u_2))] 
\ee
\be
+ [r(\gin(u_1), \gin(\hu)) - r(\fin(u_1), \fin(\hu))].
\ee
The reasoning from the previous case ($u_1 \leq \hu$) applies to the first square-bracketed term, so it is finite. For the second square-bracketed term,
\be
r(\gin(u_1),\gin(\hu)) - r(\fin(u_1), \fin(\hu)) = [\gin(u_1) - \gin(\hu)] - [\fin(u_1) - \fin(\hu)],
\ee
because $\gin(\hu) \geq 0$, $\fin(\hu) \geq 0$, and $u_1 > \hu$. We will show that 
\be
\gin(u_1) - \gin(\hu) > \fin(u_1) - \fin(\hu).
\ee
Consider the integral
\be
\int_{x = \fin(\hu)}^{\fin(u_1)} f(x) \dd{x}
\ee
By the definition of a cdf, this integral equals $u_1 - \hu$. Alternatively, perform a change of variables to $u = F(x)$, with $\dd{u} = f(x) \dd{x}$:
\be
\int_{u = \hu}^{u_1} du = u_1 - \hu.
\ee
Likewise,
\be
\int_{x = \gin(\hu)}^{\gin(u_1)} g(x) \dd{x} = u_1 - \hu.
\ee
So 
\be
\int_{x = \gin(\hu)}^{\gin(u_1)} g(x) \dd{x} = \int_{x = \fin(\hu)}^{\fin(u_1)} f(x) \dd{x}.
\ee
Therefore, the area under $g$ over the interval $[\gin(\hu), \gin(u_1)]$ is the same as the area under $f$ over the interval $[\fin(\hu), \fin(u_1)]$. Since we assume $\forall u \geq \hu: g(\gin(u)) < f(\fin(u))$, the interval $[\gin(\hu), \gin(u_1)]$ must be longer than the interval $[\fin(\hu), \fin(u_1)]$, because we must integrate over a longer interval to get the same area under a lower curve. So $\gin(u_1) - \gin(\hu) > \fin(u_1) - \fin(\hu)$, the second square-bracketed term is positive, I is bounded from below, and it can be made arbitrarily small by choosing $n$ sufficiently large. 

Now we show that II is positive. Recall that II is
\be
\int_{u_2 = \hu}^{1} \int_{u_1 = u_2}^{1} [r(\gin(u_1), \gin(u_2)) - r(\fin(u_1), \fin(u_2))] p(u_1, u_2) \dd{u_1} \dd{u_2}.
\ee
Since $\gin(\hu) \geq 0$ and $\fin(\hu) \geq 0$, $\gin(u_1)$, $\gin(u_2)$, $\fin(u_1)$, and $\fin(u_2)$ are all nonnegative in this integral, because $u_1 \geq \hu$ and $u_2 \geq \hu$. So 
\be
r(\gin(u_1), \gin(u_2)) - r(\fin(u_1), \fin(u_2)) = [\gin(u_1) - \gin(u_2)] - [\fin(u_1) - \fin(u_2)].
\ee
Since $u_1 \geq \hu$ and $u_2 \geq \hu$ in II, we can use the reasoning from before about
\be
[\gin(u_1) - \gin(\hu)] - [\fin(u_1) - \fin(\hu)],
\ee
replacing $\hu$ by $u_2$. We have equal integrals:
\be
\int_{x = \gin(u_2)}^{\gin(u_1)} g(x) \dd{x} = \int_{x = \fin(u_2)}^{\fin(u_1)} f(x) \dd{x} = u_1 - u_2.
\ee
Since $\forall u \geq \hu: g(\gin(u)) < f(\fin(u))$, the interval $[\gin(u_2), \gin(u_1)]$ must be longer than the interval $[\fin(u_2), \fin(u_1)]$. So 
\be
\forall u_1 \geq u_2 \geq \hu: [\gin(u_1) - \gin(u_2)] > [\fin(u_1) - \fin(u_2)].
\ee
As a result, II is positive. 

$\square$

%% file: appendixb_v2.tex
\noindent \begin{flushleft}
\textbf{APPENDIX B:} Proof of Theorem 4
\par\end{flushleft}

For SP expected revenue: 
\be
\mean{v_s - z_w} = \mean{v_s - z_s} + \mean{z_s - z_w}.
\ee
Note that 
\be
\mean{z_s - z_w} \leq 0,
\ee
because it is zero if $z_i$ are independent of $v_i$ and if positively correlated. (This is why we require $c\geq0$ in the theorem.). So it remains to show that
\be
\mean{\vts - \zts} \geq \mean{v_s - z_s},
\ee
where $s$ is the index of the runner-up $v_i$, and $\ts$ is the index of the runner-up $v_i - z_i$. 

Recall that
\be
z_i = \theta c v_i + (1 - \theta) y_i,
\ee
and 
\be
v_i - z_i = (1-c \theta) v_i - (1 - \theta) y_i,
\ee
where
\be
v_i \sim \norm{\mu_v}{\sigma_v},
\ee
and
\be
y_i \sim \norm{\mu_y}{\sigma_y}.
\ee
So 
\be
v_i - z_i \sim \norm{(1-c \theta) \mu_v - (1 - \theta) \mu_y}{\sqrt{(1-c \theta)^2 \sigma_v^2 + (1- \theta)^2 \sigma_y^2}}.
\ee

Since $s$ is selected without reference to $z_i$ values, 
\be
\mean{z_s | v_s} = c \theta v_s + (1 - \theta) \mu_y.
\ee
So, by the linearity of expectation,
\be
\mean{v_s - z_s} = (1-c \theta) \mean{v_s} - (1-\theta) \mu_y.
\ee
So we need to show that 
\be
\mean{\vts - \zts} \geq (1-c \theta) \mean{v_s} - (1-\theta) \mu_y.
\ee
Let
\be
\alpha_i = \frac{(v_i - z_i) + (1-\theta) \mu_y}{1 - c \theta}.
\ee
Since the mapping from $v_i - z_i$ to $\alpha_i$ preserves ordering, the runner-up $\alpha_i$ corresponds to the runner-up $v_i - z_i$. So we need to show that
\be
\mean{\alpha_{\ts}} \geq \mean{v_s}.
\ee
Note that
\be
\alpha_i \sim \norm{\mu_v}{\sqrt{\sigma_v^2 + \left( \frac{1-\theta}{1 - c \theta} \right)^2 \sigma_y^2}}.
\ee
So the distribution of $\alpha_i$ has the same mean as that of $v_i$ and a greater variance. 

We need to show that the second order statistic for a normal distribution is at least the second order statistic for another normal distribution that has the same mean and a lower variance. For $n>3$, this is true because an order statistic in the upper half of the sample has a higher mean for the higher-variance normal simply because higher variance makes higher values more likely. For $n=3$, it is true because the mean of the second order statistic is the same as the mean of the distributions: $\mu_v$. To see this, note that any set of three sample values from a distribution that is symmetric about its mean has the same likelihood as the set of reflections of the three values across the mean. But the reflected runner up is the runner up of the reflected sample set, and it is on the opposite side of the mean and equally distant from it. So the two runner ups average to the mean of the distribution.
$\square$